\documentstyle[twoside,epsf]{hsproc}
\def\piz{$\pi^0$}
\def\pim{$\pi^-$}
\def\kt{$k_T$}

\def\avkt{$\langle k_T \rangle$}
\def\mavkt{\langle k_T \rangle}
\def\avktsq{$\langle k_T^2 \rangle$}
\def\mavktsq{\langle k_T^2 \rangle}
\def\pt{$p_T$}
\def\mpt{p_T}
\def\avpt{$\langle p_T \rangle$}
\def\avptp{$\langle p_T \rangle_{pair}$}
\def\x{$x$}
\def\s{$\sqrt{s}$}
\def\DZERO{D\O}
\def\runauthor{M. Zieli\'nski}
\def\shorttitle{Effects of Parton \kt\ in High-\pt\ Particle Production}
\begin{document}
\title{Effects of Parton \kt\ in High-\pt\ Particle Production}
\author{M. Zieli\'nski 
\footnote{This work was done in collaboration with 
L. Apanasevich, C. Bal\'azs, M. Begel, C. Bromberg, T.~Ferbel, G. Ginther,
J. Huston, S. Kuhlmann, A. Maul, J. Owens, P. Slattery and W. K. Tung
\cite{ktpheno}.}
\\ \email{marek@fnal.gov}}
{University of Rochester, Rochester, NY 14627, USA}
%
\abstract{
We report on recent work 
concerning the phenomenology of initial-state parton-\kt\ effects in 
direct-photon production and related processes in hadron collisions.
After a brief summary of a \kt-smearing 
model, we present a study  of recent results on fixed-target 
and collider direct-photon production, using complementary data 
on diphoton and pion production that provide empirical guidance on the 
required amount of \kt\ broadening.  This approach provides a consistent 
description of the observed deviations of next-to-leading order 
QCD calculations relative to the inclusive direct-photon and \piz\ data.
We also  comment on the implications of these results for the 
extraction of the gluon distribution of the nucleon.}

\section*{Introduction}

	Direct-photon production has long been viewed as an ideal 
process for measuring the gluon distribution in the 
proton \cite{halzen}. The quark-gluon Compton scattering subprocess 
($gq{\rightarrow}{\gamma}q$) 
provides a large contribution to inclusive  $\gamma$ production
for which the cross sections have been calculated 
to next-to-leading order (NLO) \cite{aurenche}.
The gluon distribution in the proton is relatively
well-constrained at $x<0.1$ by deep-inelastic scattering (DIS) and 
Drell-Yan (DY) data, but less so at larger~\x.
Consequently, direct-photon data from fixed-target 
experiments can, in principle, provide an important
constraint on the gluon content at moderate to large \x. 

	However, a pattern of deviations has been observed 
between the measured 
direct-photon cross sections and NLO calculations 
\cite{cteqkt}. The discrepancy is particularly
striking in the recently published higher-statistics
data from E706 \cite{e706},
for both direct-photon and ${\pi}^{0}$ cross sections. 
The final direct-photon results from UA6 \cite{ua6} 
also exhibit evidence of similar, although smaller, discrepancies.
The suspected origin of the disagreement is 
from effects of soft-gluon radiation.
Such radiation generates transverse components 
of initial-state parton momenta, referred to in this discussion as
${k}_{T}$. (To be precise, 
\kt\ denotes the magnitude of the effective transverse momentum vector, 
${\vec k}_T$, of each of the two colliding partons.)

\begin{figure}[t]
\begin{center}
\epsfxsize=0.55\textwidth
\epsfysize=0.55\textheight
\vskip-1.5cm
\mbox{\epsfbox{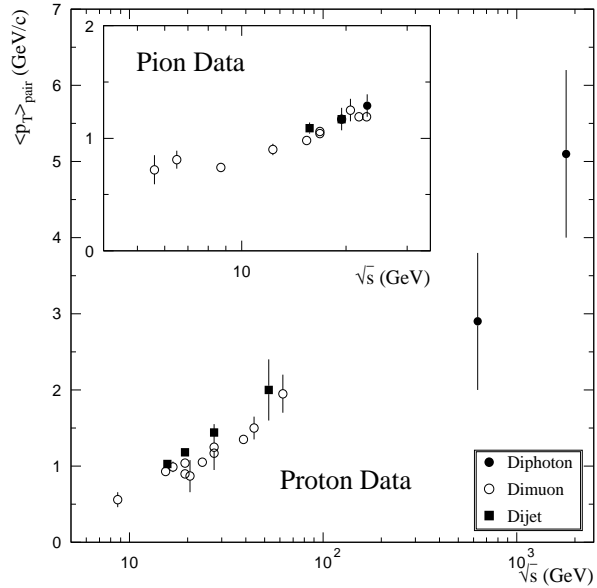}}
\vskip-2.cm
\end{center}
\caption{
\avpt\ of pairs of muons, photons, and jets produced in
hadronic collisions versus $\sqrt{s}$. 
} 
\label{fig:pairpt}
\end{figure}

Evidence of significant \kt\ has long been observed in
production of muon, photon, and jet pairs. A collection of
measurements of the average transverse momentum of the pairs
(\avptp) is displayed in Fig.~\ref{fig:pairpt}, for a wide range of 
center-of-mass energies ($\sqrt{s}$).
The values of \avptp\ are large, and increase
slowly with increasing $\sqrt{s}$.
The values of \avkt\ per parton (estimated as $\approx$\avptp/$\sqrt{2}$)
indicated by these DY, diphoton, and dijet
data, as well as the inclusive direct-photon and $\pi^0$ production data,
are too large to be interpreted as ``intrinsic" --- i.e., due only to the 
finite size of the proton. 
(From the observed data, one can infer that the average ${k}_{T}$ per parton
is about 1 GeV/$c$ at fixed-target energies, increasing 
to 3--4 GeV/$c$ at the Tevatron collider, while one would expect \avkt\ values 
on the order of 0.3--0.5 GeV/$c$ based solely on proton size.)
Perturbative QCD (PQCD) corrections at NLO level are also  insufficient
to explain the size of the observed effects, and, in fact, 
 full resummation calculations
are required to describe DY and W/Z \cite{ellis,WZ,resbos}, and diphoton 
\cite{fergani,diphoton} distributions.
Similar soft-gluon (or \kt) effects can be expected 
in all hard-scattering processes, such as the inclusive production of jets 
or direct photons~\cite{FF,cont}. 

After reviewing a phenomenological model for $k_T$ effects in 
direct-photon production 
we will discuss its applications to data, as well as 
the implications for determining the gluon distribution. 
A more detailed presentation of these results can be found in
\cite{ktpheno}.

\section*{\kt\ Smearing Model}
The Collins-Soper-Sterman resummation formalism \cite{css} provides
a rigorous basis for theoretical understanding of soft-gluon radiation
effects. Despite recent progress \cite{mang,ster}, no full treatment of
inclusive direct-photon cross sections is yet available.
In its absence, we use a PQCD-based model that incorporates transverse
kinematics of initial-state partons to study the major consequences of
\kt\ for direct-photon production
(and, by extension, for all hard-scattering processes).

In PQCD, the expression for the  leading-order (LO) cross section 
for direct-photon production at large \pt\ has the form:
$$ \sigma(h_1h_2\to \gamma X) = 
\int dx_1 dx_2 \; f_{a_1/h_1}(x_1,Q^2) \; f_{a_2/h_2}(x_2,Q^2) 
\hat{\sigma}(a_1a_2\to \gamma a_3),
$$
where $\hat{\sigma}$ is the hard-scattering matrix element,
and $f_{a_1/h_1}$ and $f_{a_2/h_2}$ are the parton distribution functions 
(pdf) for the colliding partons $a_1$ and $a_2$ in hadrons $h_1$ and $h_2$, 
respectively. To introduce $k_T$ degrees of freedom, we extend each integral 
over the parton distribution functions to the ${k}_{T}$-space, 
$$d{x} \; {f}_{a/h}({x},{Q}^2)~{\rightarrow}~d{x}{d}^2{k}_{T}
\; g({\vec k}_{T}) \; {f}_{a/h}({x},{Q}^2), \nonumber$$
and take $g({{\vec k}_{T}}$) to be a Gaussian (as justified, e.g., by E706
data on high-mass pairs),
$$g({{\vec k}_{T}})={{e^{-{k}_{T}^2/\mavktsq}}\over{{\pi}\mavktsq}}. \nonumber$$
Here, \avktsq\ is the square of the 
2-dimensional RMS width of the ${k}_{T}$ distribution for one parton 
and is related to the square of the average of the absolute value of
${\vec k}_T$ of one parton through \avktsq\ = 4$\mavkt^2/\pi$. 

Since
an exact treatment of the modified parton kinematics can be implemented in a 
Monte Carlo framework, but is more difficult in an analytic approach,
it is convenient to employ Monte Carlo techniques in evaluating 
the cross sections according to the above prescription.
In general, because the unmodified PQCD cross sections fall rapidly 
with increasing \pt, the net effect of \kt\ smearing is to increase 
the expected yield. We denote the enhancement factor as $K(\mpt)$.

	A Monte Carlo program that includes such a treatment of
${k}_{T}$ smearing, and the LO cross section
for high-\pt\ particle production, has long been available \cite{owensll}.
The program provides calculations of many experimental observables,
and can be used for 
direct photons, jets, and for single high-\pt\ particles resulting from 
jet fragmentation (such as inclusive $\pi^0$ production).
Unfortunately, no such program is available for NLO calculations, but
one can approximate the effect of \kt\ smearing by multiplying the NLO 
cross sections by the LO  ${k}_{T}$-enhancement factor.
Admittedly, this procedure involves a
risk of double-counting since some of the ${k}_{T}$-enhancement may 
already be contained in the NLO calculation. However, we expect such 
double-counting effects to be small.

A complete treatment of soft-gluon radiation in high-\pt\ production, should 
eventually predict the effective ${k}_{T}$ values expected for each 
process and \s. We will employ \avkt\ values representative of those found 
in comparisons of kinematic distributions in high-mass pair data 
with the above described model.

\section*{Applications of the \kt\ Model to Data}

The experimental consequences of \kt\ smearing are expected to
depend on the collision energy. At the Tevatron collider, the smallest photon
$p_T$ values probed by the CDF and \DZERO\ experiments are rather large
(10--15 GeV/$c$), and the \kt-enhancement factors modify only the 
very lowest end of the \pt\ spectrum, where $p_T$ is not significantly 
greater than \kt. In the E706 energy range, large \kt-effects 
can modify both the normalizations and the shapes of the cross sections as 
a function of \pt. Consequently, E706 data provide a particularly sensitive 
test of the \kt\ model. At lower fixed-target energies, the \kt\ enhancements are
expected to have less \pt\ dependence over the range of available measurements,
and can therefore be masked more easily by uncertainties in experimental 
normalizations and/or choices of theoretical scales. Nonetheless, the UA6 and WA70 
data generally support expectations from \kt\ smearing.

\subsection*{Comparisons to Tevatron Collider Data}

\begin{figure}[tbp]
\begin{center}
\vskip-1.cm
\epsfxsize=0.55\textwidth
\epsfysize=0.55\textheight
\mbox{\epsfbox{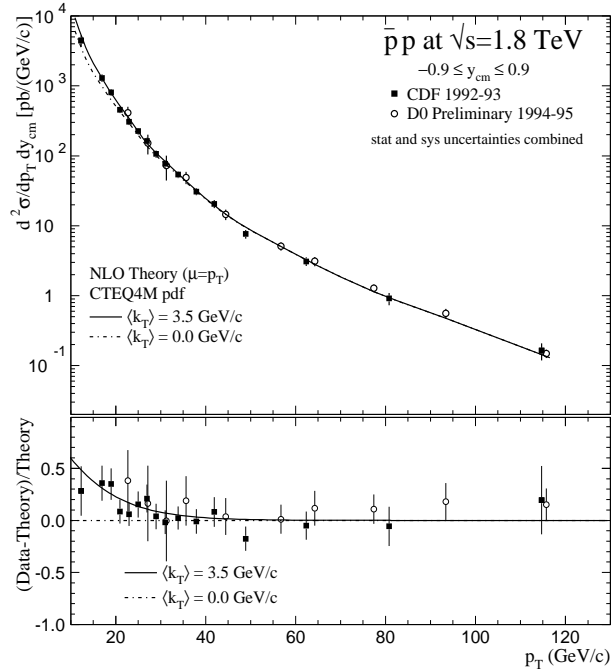}}
\end{center}
\vskip-1.5cm
\caption{
Top: The CDF and \DZERO\ isolated direct-photon cross sections, compared to
NLO theory without \kt\ (dashed) and with \kt\ enhancement for 
\avkt\ = 3.5 GeV/$c$ (solid), as a function of $p_T$.
Bottom: The quantity (Data--Theory)/Theory 
(for theory without \kt\ adjustment),
overlaid with the expected effect from ${k}_{T}$ enhancement
for \avkt\ = 3.5 GeV/$c$. 
} 
\label{fig:cdfd0}
\end{figure}

	At the Tevatron collider, the above model of soft-gluon radiation
leads to a relatively small modification of the NLO cross section.
In Fig.~\ref{fig:cdfd0} we compare the CDF and \DZERO\ isolated direct-photon 
cross sections \cite{cdfd0phot} to theoretical NLO calculations with and
without \kt\ enhancement.
In the lower part of the plot we display the quantity (Data--Theory)/Theory,
for NLO theory without the \kt-enhancement factor, and the expected effect 
from  ${k}_{T}$ for \avkt\ = 3.5 GeV/$c$. 
This is the approximate value of \avkt\ per parton  measured in diphoton 
production at the Tevatron \cite{cdfd0phot}, and one expects a similar
\avkt\ per parton for single-photon production. 
(In the diphoton process, the 4-vectors of the photons can be 
measured precisely, providing a direct determination of the transverse
momentum of the diphoton system, and thereby \avkt.)

As seen in Fig.~\ref{fig:cdfd0}, the \kt\ effect diminishes rapidly 
with increasing \pt, and is essentially negligible above  $\approx$30 GeV/$c$. 
The trend of deviations of NLO calculations from the measured inclusive 
cross sections is described reasonably well by the expected ${k}_{T}$ effect. 
Some of the observed excess can be attributed to the fragmentation effects
in the isolated direct-photon production \cite{gordon}, but this alone
cannot account for the entire deviation of the theory from data.

\subsection*{Comparisons to E706 Data}

\begin{figure}[tbp]
\begin{center}
\vskip-.8cm
\mbox{
\hskip4mm
\epsfxsize=0.44\textwidth
\epsfysize=0.4\textheight
\epsfbox{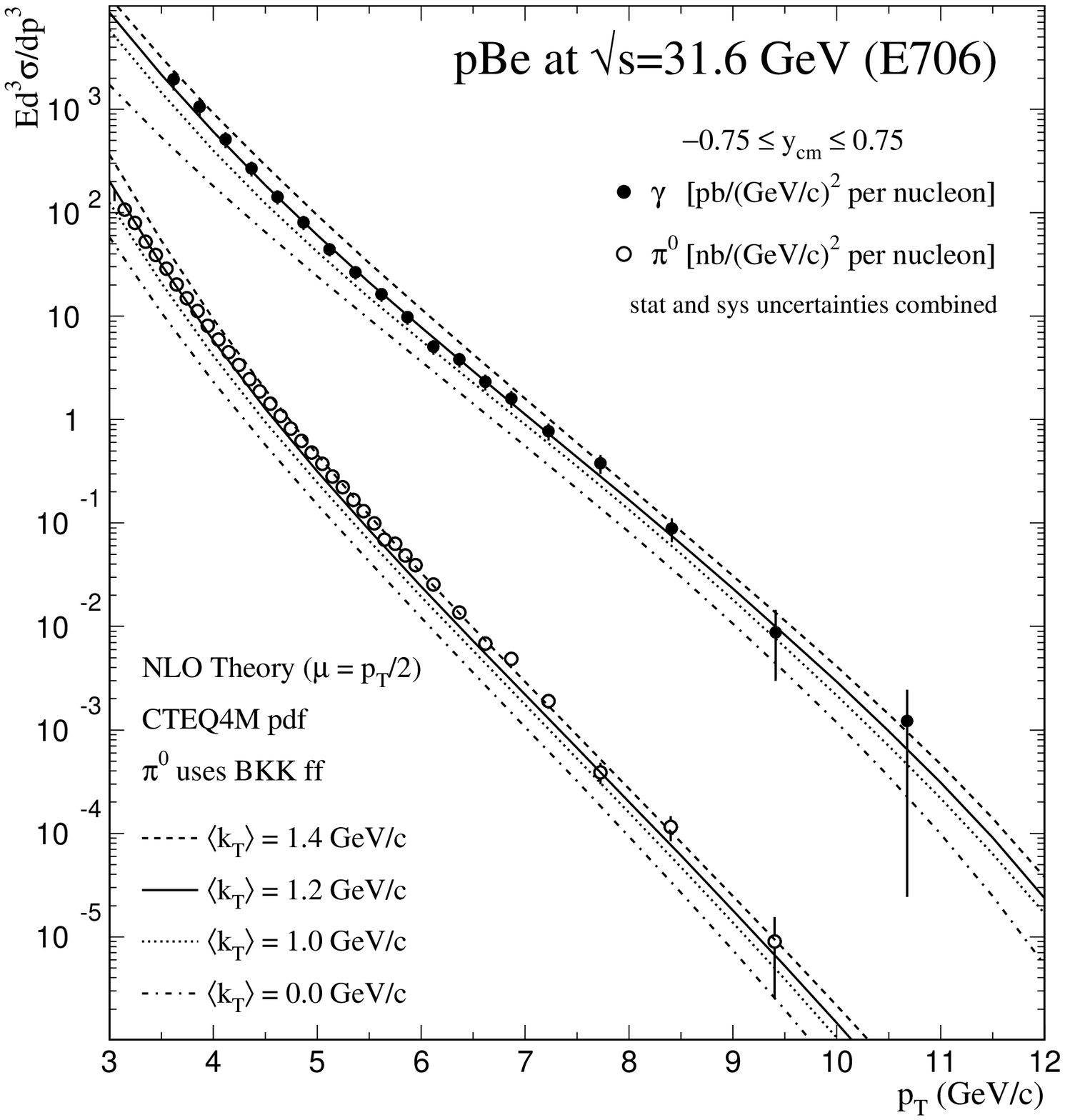}
\hskip1.cm
\epsfxsize=0.44\textwidth
\epsfysize=0.4\textheight
\epsfbox{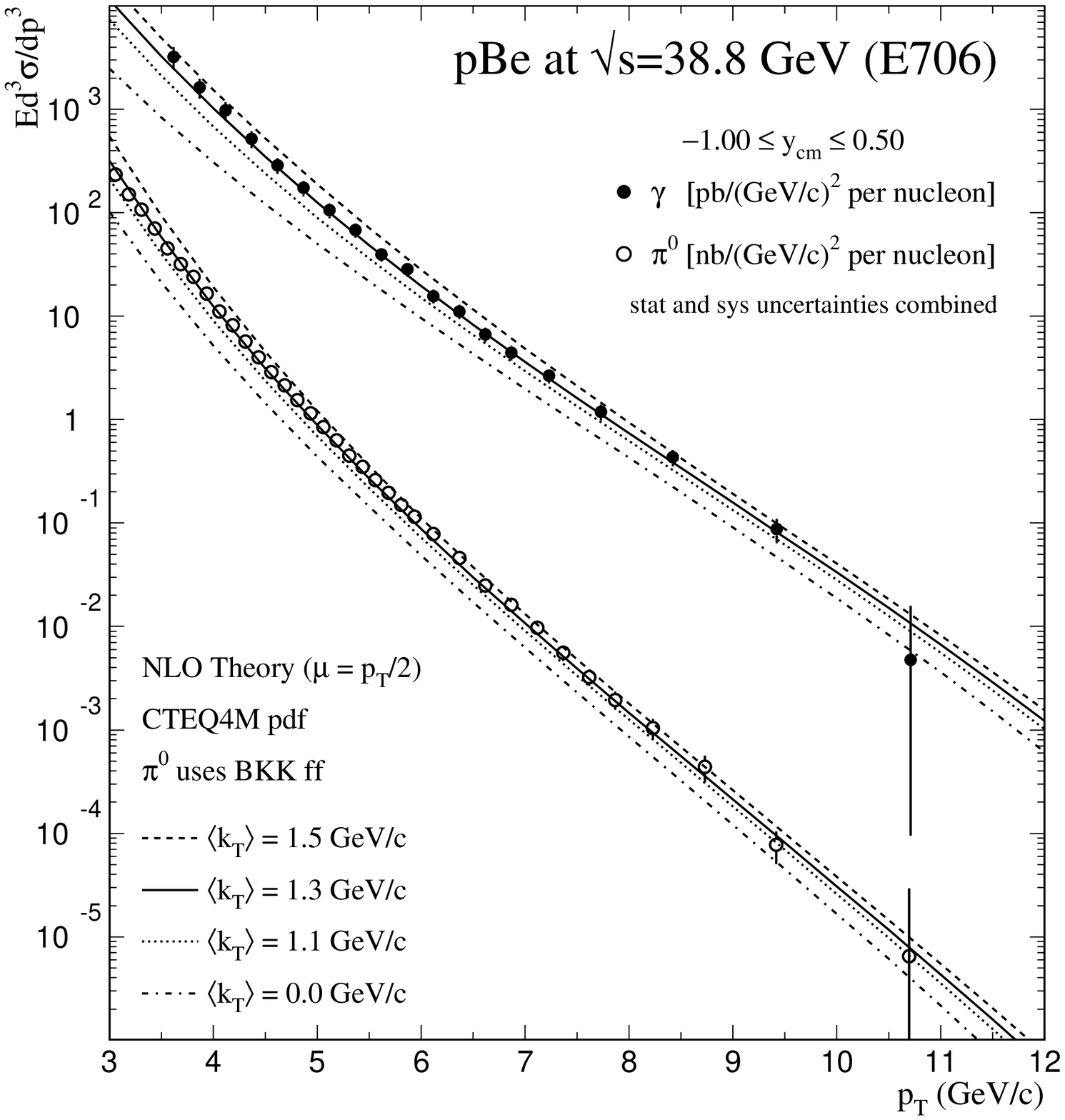}
}
\end{center}
\vskip-1.cm
\caption{
The photon and $\pi^0$ cross sections from E706 at \s\ = 31.6 GeV (left)
and 38.8 GeV (right) compared to
${k}_{T}$-enhanced NLO calculations.
} 
\label{fig:xs530}
\end{figure}

The conventional (\avkt=0) NLO calculations yield cross sections that are 
signficantly below the E706 direct-photon and $\pi^0$ measurements \cite{e706} 
(see Fig.~\ref{fig:xs530}).
No choices of current parton distributions, or conventional
PQCD scales, provide an adequate description of the data
(for the presented comparisons all QCD scales have been set to \pt/2). 
The previously described ${k}_{T}$-enhancement algorithm was used to 
incorporate the effects of soft-gluon radiation in the 
calculated yields. That is, the theory results plotted in the figures 
represent the NLO calculations multiplied by ${k}_{T}$-enhancement factors 
${K}$(${p}_{T}$).

As seen in Fig.~\ref{fig:xs530},
the NLO theory, when supplemented with appropriate ${k}_{T}$
enhancements, is successful in describing both the shape and normalization
of the E706 direct-photon cross sections at both \s\ = 31.6 GeV and 38.8 GeV.
As expected, the ${k}_{T}$-enhancement factors affect the normalization of 
the cross sections, as well as the shapes of the \pt\ distributions. 
The values of \avkt\ = 1.2 GeV/$c$ at $\sqrt{s}$ = 31.6 GeV, and 1.3 GeV/$c$ 
at $\sqrt{s}$ = 38.8 GeV, provide good representations of the incident-proton 
data. Both \avkt\ values are consistent with those emerging  from a comparison 
of the same PQCD Monte Carlo with E706 data on the production of high-mass 
$\pi^0\pi^0$, $\gamma\pi^0$, and $\gamma\gamma$ pairs \cite{e706}.
Similar conclusions are reached from comparisons between calculations and 
E706 data for $\pi^-$Be interactions at \s\ = 31.1 GeV (not shown).

\begin{figure}[tb]
\begin{center}
\vskip-.5cm
\mbox{
\hskip4mm
\epsfxsize=0.44\textwidth
\epsfysize=0.4\textheight
\epsfbox{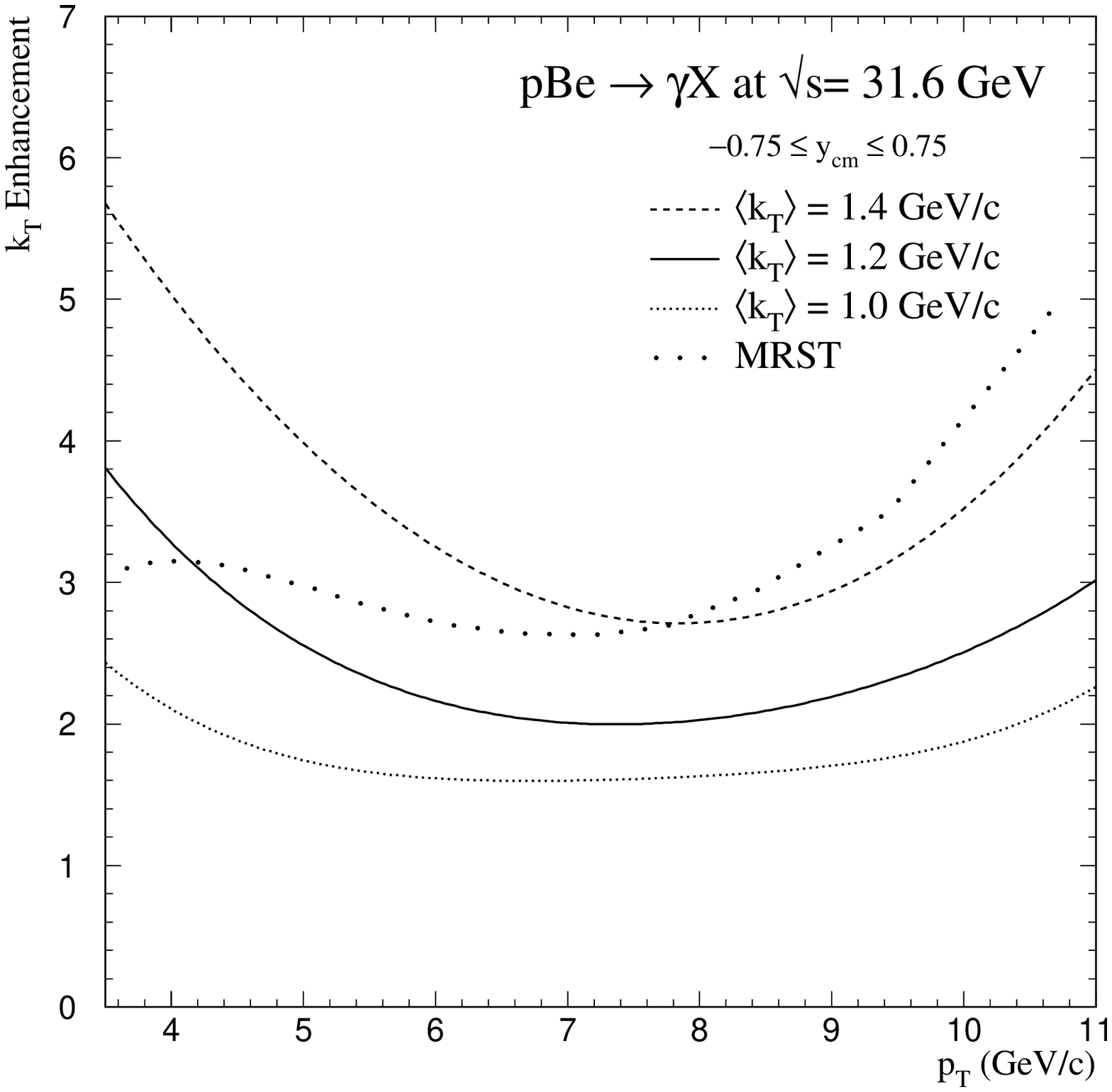}
\hskip1cm
\epsfxsize=0.44\textwidth
\epsfysize=0.4\textheight
\epsfbox{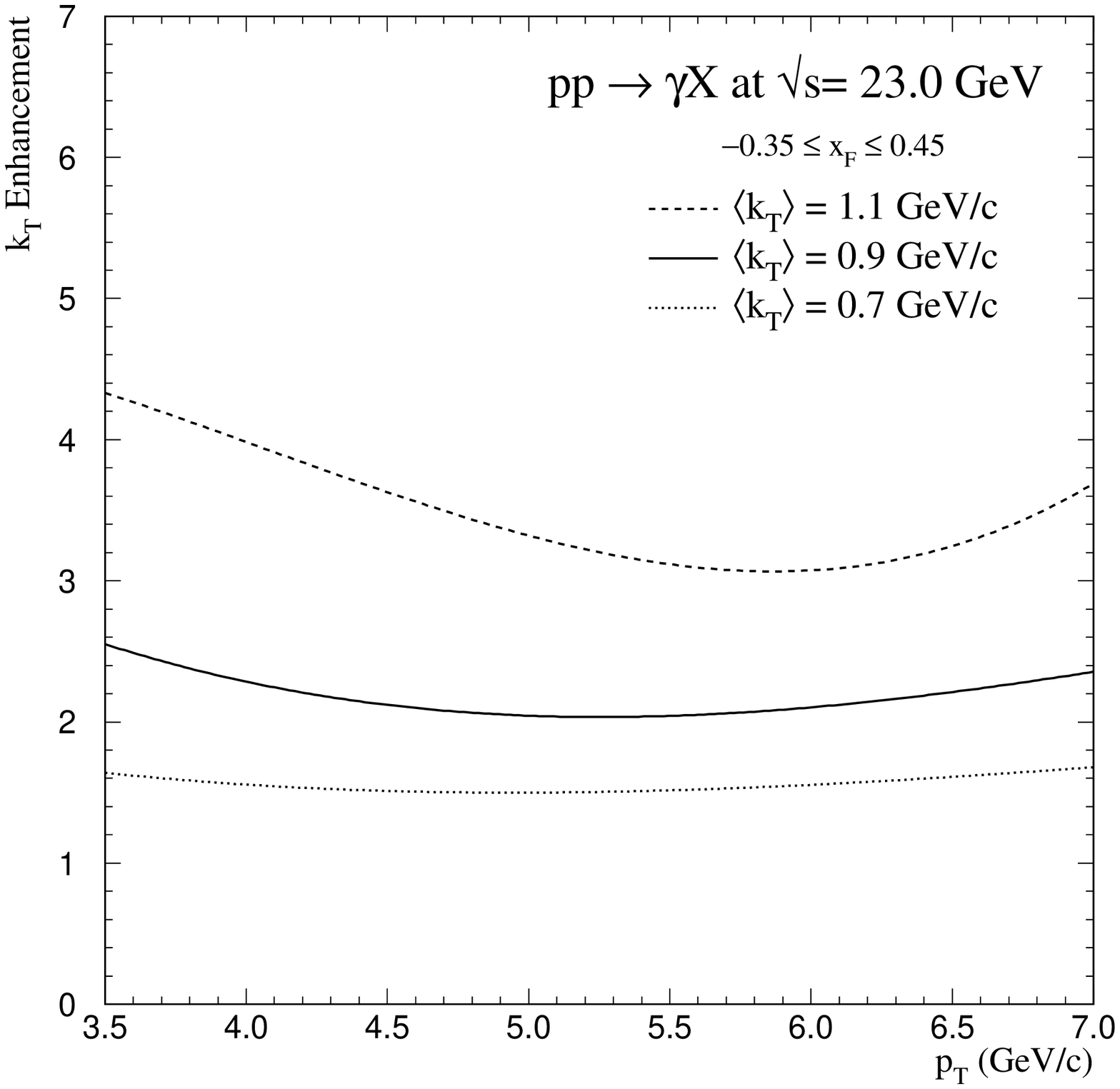}
}
\vskip-1.2cm
\end{center}
\caption{
Left: The variation of ${k}_{T}$ enhancements, $K(\mpt)$,
relevant to the E706 direct-photon 
data for protons at \s\ = 31.6 GeV, for different values of \avkt.
$K(\mpt)$ used by the MRST group {\protect \cite{mrst}}
is also shown.
Right: The same for \s\ = 23.0 GeV (relevant to WA70 data).
} 
\label{fig:kfac530}
\end{figure}

For comparison, results of calculations using \avkt\ values $\pm$0.2 GeV/$c$
relative to the central values are also shown in Fig.~\ref{fig:xs530}. These
can be taken as an indication of uncertainties on \avkt, on the
basis of several  considerations. These include: (i) the range of \kt\ values 
inferred from different distributions in the E706 high-mass pair data;
(ii) differences observed between photon and $\pi^0$ results;
(iii) comparisons of \avkt\ values required in inclusive cross sections with
those representing the properties of massive pairs at E706 and 
WA70/UA6 energies, and (iv) the differences between dimuon, diphoton, 
and dijet values of \avptp\ seen in Fig.~\ref{fig:pairpt}. Examples of
enhancement factors $K(\mpt)$ for fixed-target 
experiments are displayed in Fig.~\ref{fig:kfac530}.

	It is interesting to note that, for the fixed-target energy range, 
the ${k}_{T}$ enhancement increases at the  highest values of ${p}_{T}$.
The shape of $K(\mpt)$ can be understood through the following argument.
At the low ${p}_{T}$ end of the measurements, a \avkt\  of $\approx$1 GeV/$c$ 
is non-negligible
in comparison to the \pt\ in the hard-scattering, and the addition of
${k}_{T}$ smearing therefore increases the size of the cross section 
(and steepens the slope). At highest ${p}_{T}$ (corresponding to large \x), 
the unmodified NLO cross section becomes 
increasingly steep (due to the rapid fall in parton densities),
and hence the effect of ${k}_{T}$ smearing again becomes larger.

	NLO calculations for $\pi^0$ production have a greater theoretical 
uncertainty than those for direct-photon production since $\pi^0$ production 
involves parton fragmentation. However, the ${k}_{T}$ effects 
in $\pi^0$ production can be expected to be similar to those 
observed in direct-photon production, and the $\pi^0$ data can be used 
to extend tests of the consequences of \kt\ smearing. Figure~\ref{fig:xs530}
also shows comparisons between NLO calculations \cite{aversa} and $\pi^0$ 
production data from E706, using BKK fragmentation functions (ff) \cite{BKK}.
The previously described Monte Carlo program was employed to generate 
${k}_{T}$-enhancement factors for $\pi^0$ cross sections, and \avkt\ per 
parton values similar to those that provided good agreement 
for direct-photon data also provide a reasonable description of $\pi^0$ 
data. For $\pi^0$ production, an additional smearing 
of the transverse momentum expected from jet fragmentation
has also been taken into account.

\subsection*{Comparisons to WA70 and UA6 Data}

Both WA70 and UA6 have measured direct-photon production with 
good statistics, and their data have been
included in recent global fits to parton distributions.
WA70 measured direct-photon and $\pi^0$ production in 
$pp$ and $\pi^-p$ collisions at $\sqrt{s}$ = 23.0 GeV \cite{wa70phot}, and
UA6 has recently published \cite{ua6} their final results (with 
substantially reduced uncertainties) for direct-photon production
in $pp$ and $\overline{p}p$ collisions at \s\ = 24.3 GeV.
These center of mass energies are smaller than 
those of E706, and the \avkt\ values are therefore expected to be smaller
(perhaps of order 0.7--0.9 GeV/$c$, based on Fig.~\ref{fig:pairpt}). WA70 has 
compared kinematic distributions observed in diphoton events (for $\pi^- p$ 
interactions) to NLO predictions, and has found that smearing the NLO theory 
with an additional \avkt\ of 0.9$\pm$0.2 GeV/$c$ provides agreement 
with their data \cite{wa70diphot}.
We therefore use this \avkt\ as the central value for the
\kt-enhancement factors for both experiments, and vary the \avkt\ by
$\pm0.2$ GeV/$c$, as was done with E706. 
Over the narrower ${p}_{T}$ range of WA70 and UA6 measurements, the effect 
of ${k}_{T}$ is essentially to produce a shift in normalization,
as illustrated on the right side of Fig.~\ref{fig:kfac530}.

\begin{figure}[tbp]
\begin{center}
\vskip-.8cm
\mbox{
\hskip4mm
\epsfxsize=0.44\textwidth
\epsfysize=0.4\textheight
\epsfbox{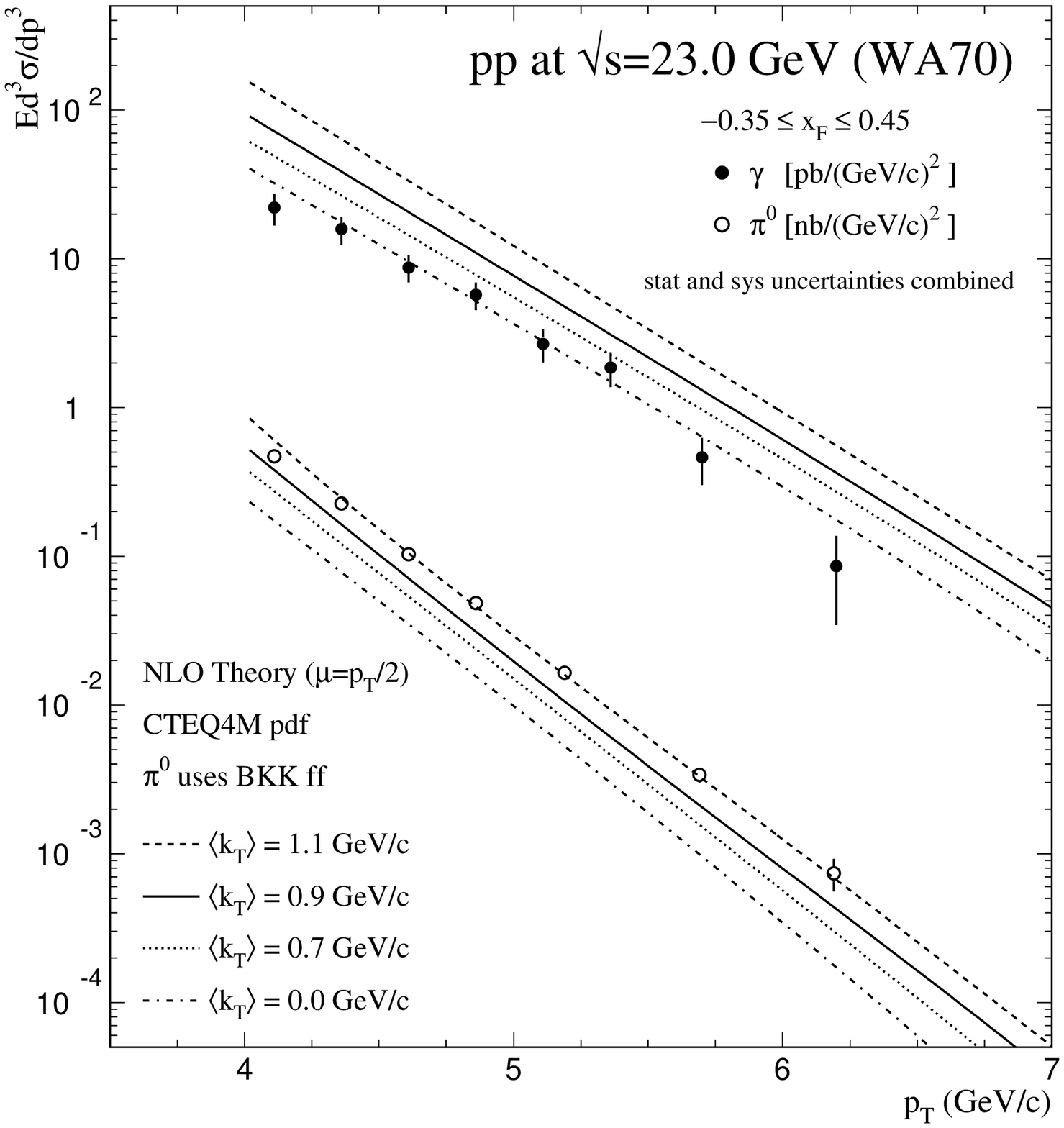}
\hskip1cm
\epsfxsize=0.44\textwidth
\epsfysize=0.4\textheight
\epsfbox{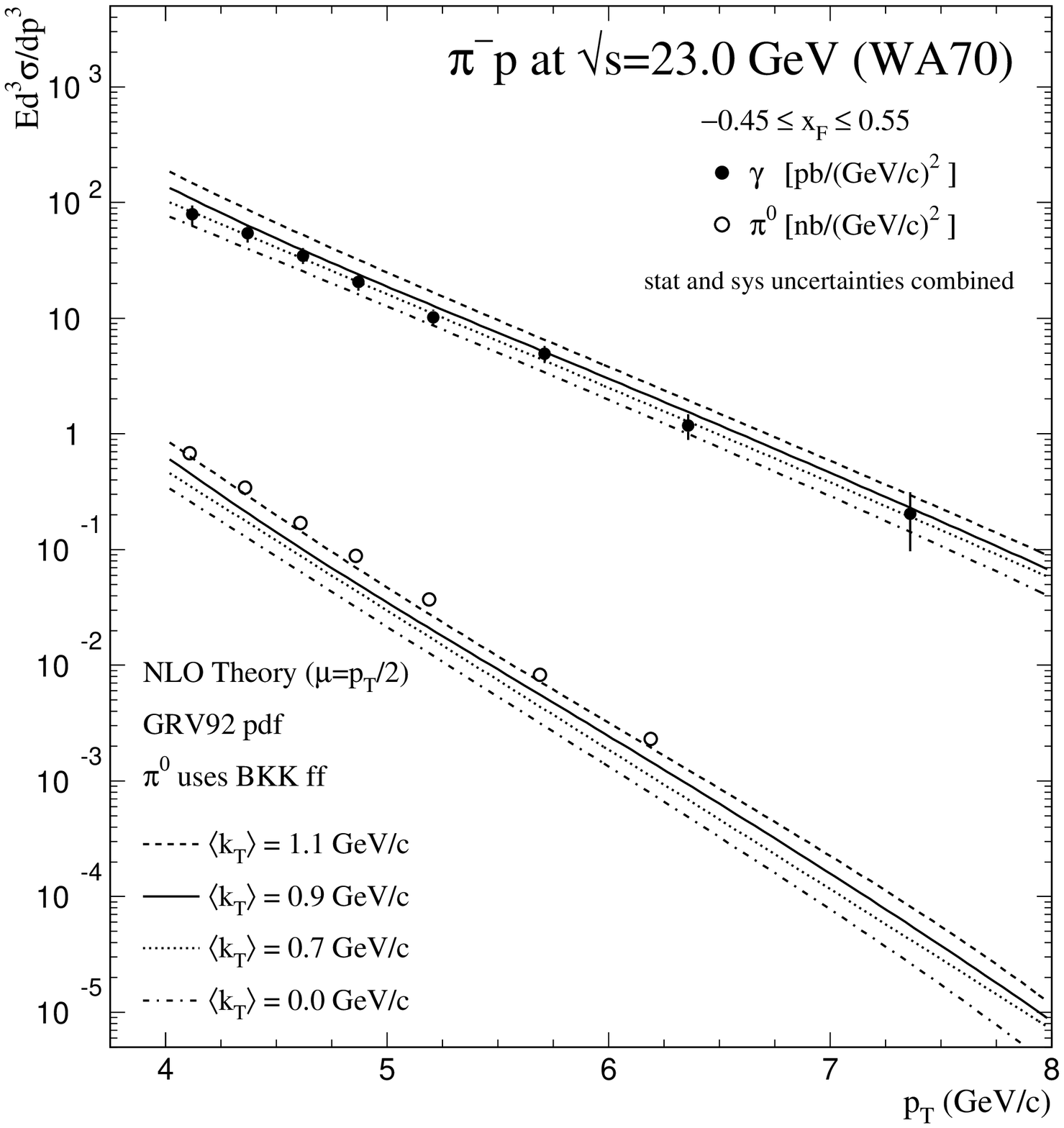}
}
\end{center}
\vskip-1cm
\caption{
The photon  and ${\pi}^{0}$ cross sections from WA70 at 
\s\ = 23.0 GeV for incident protons (left) and \pim\ (right),
compared to ${k}_{T}$-enhanced 
NLO calculations.
} 
\label{fig:xs280p}
\end{figure}

Comparisons of the WA70 direct-photon and ${\pi}^{0}$ 
cross sections with the ${k}_{T}$-enhanced NLO calculations are
shown in Fig.~\ref{fig:xs280p} (using QCD scales of ${p}_{T}$/2).
The ${\pi}^{0}$ cross sections both for incident proton and ${\pi}^{-}$ 
beams, and the photon data from incident ${\pi}^{-}$ beam,
all lie above the NLO calculations for \avkt = 0, and are in better agreement
with the ${k}_{T}$-enhanced calculations; only the photon cross section for
incident protons seems not to require a \kt\ correction.

\begin{figure}[tbp]
\begin{center}
\vskip-.7cm
\mbox{
\hskip4mm
\epsfxsize=0.44\textwidth
\epsfysize=0.4\textheight
\epsfbox{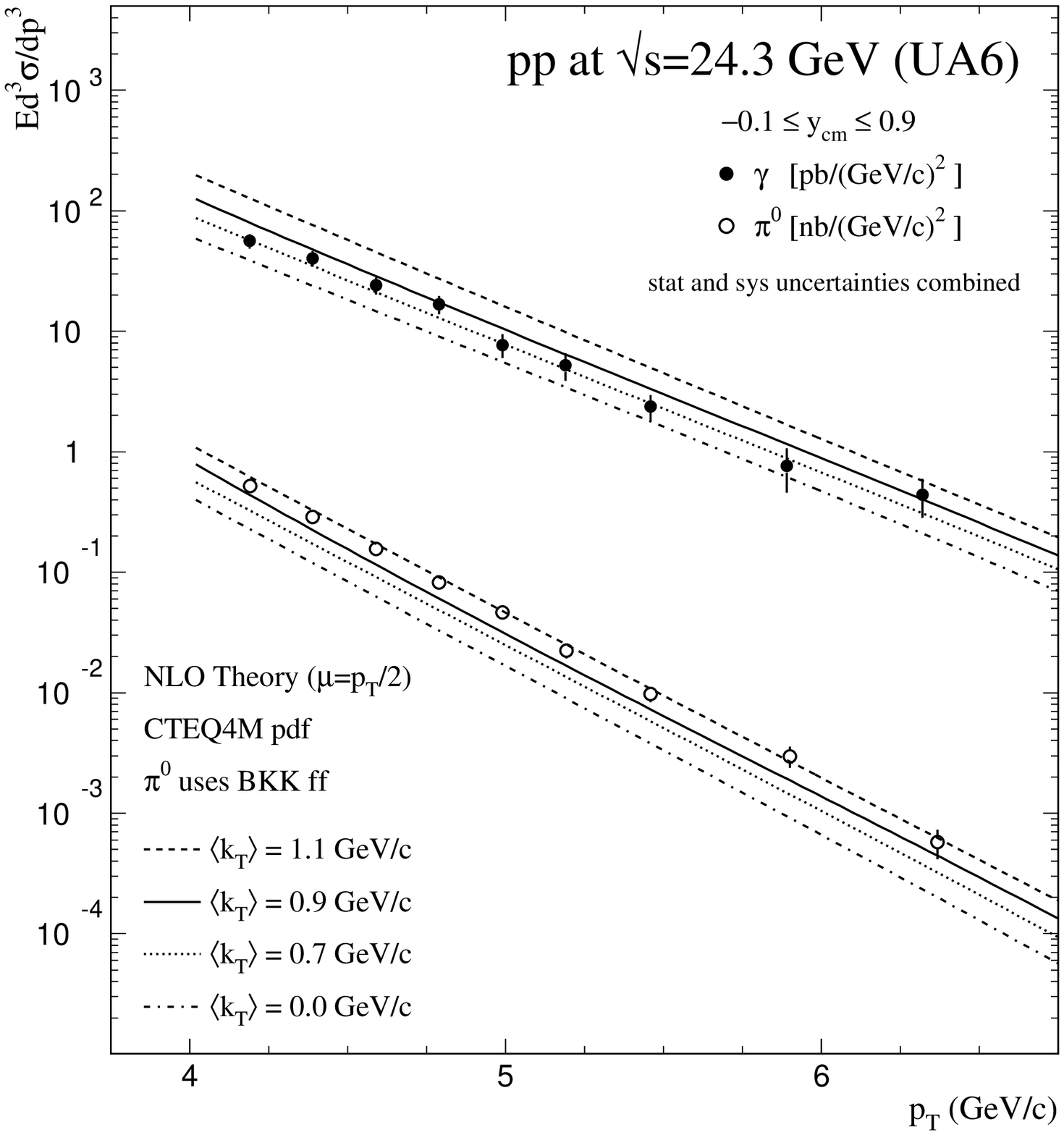}
\hskip1cm
\epsfxsize=0.44\textwidth
\epsfysize=0.4\textheight
\epsfbox{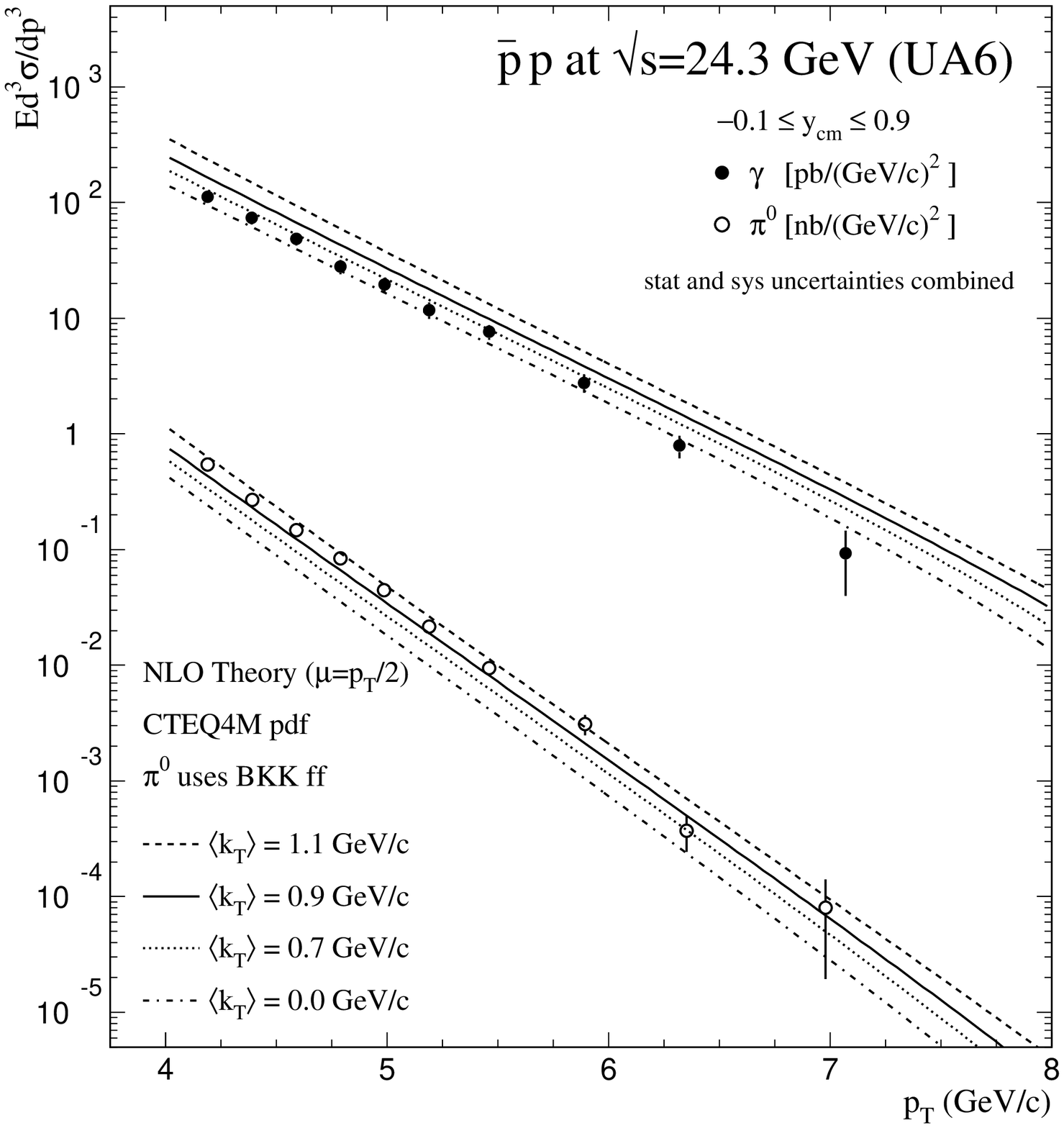}
}
\end{center}
\vskip-1cm
\caption{
The photon and $\pi^0$ cross sections from UA6 at \s\ = 24.3 GeV for 
incident protons (left) and antiprotons (right),
compared to ${k}_{T}$-enhanced NLO calculations.
} 
\label{fig:xs315p}
\end{figure}

The photon and \piz\ cross sections from UA6  for $pp$ and $\overline{p}p$ 
scattering are shown in Fig.~\ref{fig:xs315p}.
The photon cross section for $pp$ interactions lies clearly above the 
NLO calculation for \avkt\ = 0, but is consistent with \kt-adjusted
calculations for \avkt\ in the range of 0.7--0.9 GeV/$c$. The result
for $\overline{p}p$ interactions is also above the unmodified NLO calculation,
but requires a smaller value of \avkt. We note that the dominant
production mechanisms for the two processes are different: quark-gluon
Compton scattering dominates for $pp$, and  $\overline{q}q$
annihilation for $\overline{p}p$ at the UA6 energy.
As in the case of E706 and WA70, the UA6 $\pi^0$ cross sections
are higher than the NLO calculation without \kt, and can be described 
much better by introducing \kt\ enhancement.

\section*{Impact on the Gluon Distribution}

\begin{figure}[tb]
\begin{center}
\vskip-1.5cm
\epsfxsize=0.6\textwidth
\epsfysize=0.6\textheight
\mbox{\epsfbox{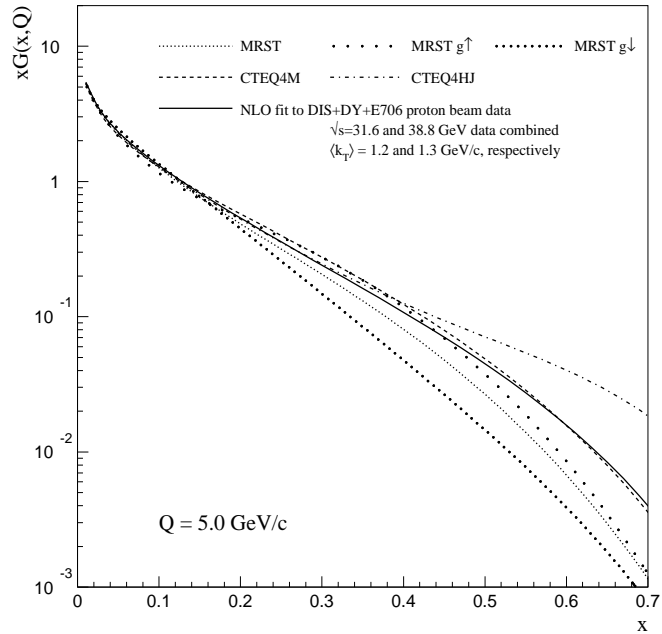}}
\vskip-2.1cm
\end{center}
\caption{
A comparison of the  CTEQ4M, MRST, and CTEQ4HJ gluons, 
and the gluon distribution derived from fits that use E706 data.
The $g\!\!\uparrow$ and $g\!\!\downarrow$ gluon densities correspond to
the maximum variation in \avkt\ that MRST allowed in their fits.
} 
\label{fig:gluons}
\end{figure}

It is now generally accepted that the uncertainty in the gluon distribution 
at large \x\ is still quite large. Thus, it would appear important to incorporate
further constraints on the gluon, especially from direct-photon data.
To investigate the impact of \kt\ effects on determinations of the gluon 
distribution, we have included the E706 direct-photon cross sections for 
incident protons, along with the DIS and DY data that were used in determining 
the CTEQ4M pdfs, in a global fit to the parton distribution functions.
The CTEQ fitting program was employed to obtain these 
results \cite{cteqfit}, using the NLO PQCD calculations for direct-photon
cross sections, adjusted by the ${k}_{T}$-enhancement factors.
However, the WA70, UA6, CDF, and \DZERO\ data were excluded from
this particular fit.
The resulting gluon distribution, shown in Fig.~\ref{fig:gluons}, is 
similar to CTEQ4M, as might have been expected, since
the ${k}_{T}$-enhanced NLO cross sections using CTEQ4M 
provide a reasonable description of the data
shown  in Fig.~\ref{fig:xs530}.

The new MRST gluon distribution \cite{mrst} (also shown in Fig.~\ref{fig:gluons}) 
is significantly lower than CTEQ4M at large \x. 
While the MRST fit employs \kt\ enhancements
(obtained using an analytic integration technique), it attempts to
accommodate the WA70 incident-proton direct-photon data, 
which does not exhibit an obvious \kt\ effect. In addition, the MRST 
\kt-enhancements are larger than ours at large \pt\ 
(see Fig.~\ref{fig:kfac530}), resulting in a smaller gluon at large 
\x\ \cite{avkt}.
In contrast, the CTEQ4HJ gluon distribution \cite{cteq4hj}, designed
to improve the description of the high-${p}_{T}$ jet data from CDF
in Run IA, is much larger than CTEQ4M in the
same \x\ range. This spread of the solutions for the gluon
distribution at large \x\ is uncomfortably large, and additional theoretical
effort is warranted to properly incorporate the 
available direct-photon data in the pdf fits.

\section*{Conclusions} 

We have described a phenomenological model in which 
\avkt\ values used in the calculations of \kt-enhancements 
are derived from data.  The results are remarkably successful
in reconciling the data and theoretical calculations for a broad range
of energies. The \kt-enhancements improve the agreement
of PQCD calculations with E706, UA6, and $\pi^-$ beam WA70 direct-photon 
cross sections over the full \pt\ range of measurements, as well as with the
low-$p_T$ end of CDF and \DZERO\ results. All discussed fixed-target $\pi^0$
measurements also agree much better with such \kt-enhanced calculations.

A definitive conclusion regarding the quantitative role of ${k}_{T}$ effects 
in hard scattering awaits a more rigorous theoretical 
treatment of soft-gluon radiation. Such theoretical progress
is crucial for a more reliable determination of the 
gluon distribution, especially in the large-\x\ region, where significant 
uncertainties remain.



\begin{thebibliography}{99}
%
%
%
\bibitem{ktpheno}
L. Apanasevich {\it et al.}: hep-ph/9808467, submitted to {\it Phys. Rev. D};
\bibitem{halzen}
\refer{F. Halzen and D. Scott} {Phys. Rev. D} {21} {1980} {1320}
\bibitem{aurenche}
P. Aurenche {\it et al.}: {\it Phys. Lett.}~{\bf 140B} (1984) 87; 
{\it Nucl. Phys. B}~{\bf 297} (1988) 661;
\bibitem{cteqkt}
\refer{J. Huston {\it et al.}}{ Phys.\ Rev. D} {51} {1995} {6139}
\bibitem{e706}
\refer{L. Apanasevich {\it et al.}} {Phys. Rev. Lett.}{81}{1998}{2642}
\bibitem{ua6}
\refer{G. Ballocchi {\it et al.}}{Phys. Lett. B}{436}{1998}{222}
\bibitem{ellis}
\refer{G. Altarelli, R. K. Ellis, M. Greco, G. Martinelli}
{Nucl. Phys. B} {246} {1984} {12}
\bibitem{WZ}
\refer{C. Bal{\'{a}}zs, C.-P. Yuan, J.-W. Qiu}
{Phys.~Lett. B}{355}{1995}{548}
\bibitem{resbos}
C. Bal{\'{a}}zs and C.-P. Yuan: {\it Phys.~Rev. D}~{\bf 56}~(1997)~5558,
and references therein;
\bibitem{fergani}
\refer{P. Chiappetta, R. Fergani, J. Ph. Guillet}{Phys. Lett. B}{348}{1995}{646}
\bibitem{diphoton}   
\refer{C. Bal\'azs, E. Berger, S. Mrenna, C.-P. Yuan}{Phys.~Rev. D}{57}{1998}{6934}
\bibitem{FF}
\refer{R. Feynman, R. Field, G. Fox}{Phys. Rev. D}{18}{1978}{3320}
\bibitem{cont}
\refer{A. P. Contogouris {\it et al.}}{Nucl. Phys. B}{179}{1981}{461}
{\it Phys. Rev. D} {\bf 32} {(1985)} {1134};
\bibitem{css}
\refer{J.~Collins, D.~Soper, G.~Sterman}
{Phys. Lett. B}{109}{1982}{388}
{\it ibid.\,B}\,{\bf 126}\,(1983)\,275;
{\it Nucl.~Phys. B}~{\bf 223}~(1983) 381;
{\it ibid. B}~{\bf 250}~(1985) 199;
\bibitem{mang}
\refer{S. Catani, M. Mangano, P. Nason}{J. High Energy Phys.}{07}{1998}{024}
\bibitem{ster}
E. Laenen, G. Oderda, G. Sterman: {hep-ph/9806467}; \\
\refer{N. Kidonakis, G. Oderda, G. Sterman}{Nucl. Phys. B}{525}{1998}{299}
{hep-ph/9803241}; 
\bibitem{owensll}
\refer{J. Owens}{Rev. Mod. Phys.}{59}{1987}{465}
\bibitem{cdfd0phot}
See, e. g., S. Linn at the ICHEP conference in Vancouver, July 1998; 
\bibitem{gordon}
\refer{L. E. Gordon}{Nucl. Phys. B}{501}{1997}{175}
\bibitem{mrst}
\refer{A. D. Martin, R. G. Roberts, W. J. Stirling, R. S. Thorne}{Eur. Phys. J. C}{4}{1998}{463}
\bibitem{aversa}
\refer{F. Aversa {\it et al.}}{Nucl. Phys. B}{237}{1989}{105}
\bibitem{BKK}
\refer{J. Binnewies, B. A. Kniehl, G. Kramer}{Phys. Rev. D}{52}{1995}{4947}
\bibitem{wa70phot}
\refer{M. Bonesini {\it et al.}}{Z. Phys. C}{38}{1988}{371}
{\it ibid.\,C\,}~{\bf 37}\,(1988)\,535;
{\it ibid.\,C }{\bf 37}\,(1987)\,39;
\bibitem{wa70diphot}
\refer{E. Bonvin {\it et al.}}{Phys. Lett. B}{236}{1990}{523}
{\it Z. Phys. C}~{41} (1989) 591;
\bibitem{cteqfit}
\refer{H. L. Lai {\it et al.}}{Phys. Rev. D}{55}{1997}{1280}
\bibitem{avkt}
See \cite{ktpheno} for a discussion of \avkt\ values used in our
approach and by MRST;
\bibitem{cteq4hj}
\refer{J. Huston {\it et al.}}{Phys. Rev. Lett.}{77}{1996}{444}
%
\end{thebibliography}
\end{document}